\documentclass[aps,showpacs,superscriptaddress,twocolumn]{revtex4}

\usepackage{amsfonts}
\usepackage{amsmath}
\usepackage{amsthm}
\usepackage{amscd}
\usepackage{amssymb}
\usepackage{amsxtra}
\usepackage{bm}           
\usepackage[all]{xy}
\usepackage{bbm}
\usepackage{exscale}
\usepackage{epic,rotating}
\usepackage{subfigure}
\usepackage{graphics,color,dsfont}
\usepackage{latexsym}
\usepackage{enumerate}
\usepackage{dcolumn}      
\usepackage{pstricks,pst-grad,fancybox,graphics}
\usepackage[scanall]{psfrag}
\usepackage[dvips]{epsfig}

\newcommand{\Akommut}[2]{\ensuremath{\{ #1,#2 \}}}

\newcommand{\varce}[1]{\ensuremath{\delta^2(#1)}}

\newcommand{\bra}[1]{\ensuremath{\langle#1|}}

\newcommand{\ket}[1]{\ensuremath{|#1\rangle}}

\newcommand{\ketbra}[1]{\ensuremath{| #1 \rangle \langle #1 |}}

\newcommand{\eins}{\ensuremath{\mathbbm 1}}
\newcommand{\tilt}{\ensuremath{\tilde{t}}}

\newcommand{\HH}{\ensuremath{\mathcal{H}}}
\newcommand{\fa}{\ensuremath{\mathfrak{a}}}
\newcommand{\fe}{\ensuremath{\mathfrak{e}}}
\newcommand{\EE}{\ensuremath{\mathcal{E}}}
\newcommand{\BB}{\ensuremath{\mathcal{B}}}
\newcommand{\PP}{\ensuremath{\mathcal{P}}}
\newcommand{\blockmatrix}[4]{\ensuremath\left(\begin{array}{cc}#1 & #2 \\ #3 & #4 \end{array}\right)}
\newcommand{\bear}{\begin{eqnarray}}
\newcommand{\eear}{\end{eqnarray}}
\newcommand{\bearn}{\begin{eqnarray*}}
\newcommand{\eearn}{\end{eqnarray*}}
\newcommand{\kommentar}[1]{}

\newcommand{\mean}[1]{\ensuremath{\langle #1 \rangle}}

\newcommand{\tr}[1]{\ensuremath{\mbox{Tr}( #1 )}}

\newcommand{\bc}{\begin{center}}
\newcommand{\ec}{\end{center}}

\renewcommand{\qed}{\ensuremath{\hfill \blacksquare}\medskip}

\newtheorem{thm}{Theorem}
\newtheorem{lem}[thm]{Lemma}

\newtheorem{proposition}[thm]{Proposition}
\newtheorem{defn}[thm]{Definition}
\renewcommand{\vr}{\ensuremath{\varrho}}
\def\bi#1\ei {\begin{itemize}#1\end{itemize}}
\def\bea#1\eea {\begin{align}#1\end{align}}
\def\bean#1\eean {\begin{align*}#1\end{align*}}
\def\ben#1\een {\begin{equation*}#1\end{equation*}}
\def\be#1\ee {\begin{equation}#1\end{equation}}
\def\bes#1\ees {\begin{equation}\begin{split}#1\end{split}\end{equation}}

\newcommand{\halbe}{\frac{1}{2}}

\begin{document}

\title{Quantifying entanglement with covariance matrices}

\author{Oleg Gittsovich}

\affiliation{Institut f\"ur Quantenoptik und Quanteninformation,
~\"Osterreichische Akademie der Wissenschaften, Otto-Hittmair-Platz 1,
6020 Innsbruck, Austria}

\author{Otfried G\"uhne}
\affiliation{Institut f\"ur Quantenoptik und Quanteninformation,
~\"Osterreichische Akademie der Wissenschaften, Otto-Hittmair-Platz 1,
6020 Innsbruck, Austria}

\affiliation{Institut f\"ur Theoretische Physik,
Universit\"at Innsbruck, Technikerstra{\ss}e 25,
6020 Innsbruck, Austria}

\begin{abstract}
Covariance matrices are a useful tool to investigate correlations and 
entanglement in quantum systems. They are widely used in continuous 
variable systems, but recently also for finite dimensional systems 
powerful entanglement criteria in terms of covariance matrices have 
been derived. We show how these results can be used for the quantification 
of entanglement in bipartite systems. To that aim we introduce an 
entanglement parameter that quantifies the violation of 
the covariance matrix criterion and can be used to 
give a lower bounds on the concurrence. These lower bounds are 
easily computable and give entanglement estimates for many 
weakly entangled states.
\end{abstract}


\date{\today}

\maketitle

\section{Introduction}

Entanglement is a central resource in quantum information processing
and many works are devoted to its characterization 
\cite{hororeview,witreview,plenioreview}. One 
line of research is the derivation of entanglement criteria which
should detect the entanglement also of weakly entangled states. A 
different line of research tries to quantify the entanglement via 
so-called entanglement measures \cite{plenioreview}. 
As most entanglement measures 
are defined via complex optimization procedures, they are often 
difficult to compute and therefore one tries to give at least
lower bounds on them \cite{chenbounds,otherconvexbounds,generallowerbounds}.

Covariance matrices (CMs) of local observables are a widely used 
tool to study correlations in continuous variable systems, such as 
coupled harmonic oscillators or modes of light \cite{loockreview}. 
Moreover, many entanglement criteria for these systems are formulated 
as conditions on CMs \cite{cmgauss}. Recently, it has been shown that CMs 
are also a useful 
tool for the investigation of entanglement in discrete systems, such 
as polarized photons or trapped ions \cite{oldprl,wir,wirPRA}. 
Indeed, in Refs.~\cite{wir,wirPRA} a so-called 
covariance matrix criterion (CMC) has been established, which allows to 
detect many weakly entangled states, which are not detected by other 
criteria. 

In this paper we show that CMs can be used not only to detect but also 
to quantify entanglement in discrete composite quantum systems. To that 
aim, we define an entanglement parameter $\EE(\vr)$, that quantifies 
the violation of the CMC for discrete systems. Our construction is 
inspired by a similar definition of an entanglement parameter for 
continuous variable systems in Ref.~\cite{giedkecirac}. While it remains 
unclear to which extent $\EE(\vr)$  is a direct entanglement measure, 
we will see, however, that $\EE(\vr)$ gives a lower bound on the concurrence, 
which is a widely used entanglement monotone.

In detail, our paper is organized as follows: In Section II we recall 
the CMC and define our entanglement parameter $\EE$. We also demonstrate 
that existing results about the CMC give directly lower bounds on $\EE$. 
In Section~III we consider general properties of the the parameter $\EE$.
We show that is is convex and invariant under local rotations, but we 
present an example, where $\EE$ increases on average under LOCC. The physical 
reason behind this example is the fact that there are entangled states, 
which cannot be detected by the CMC, however, they are detected after
suitable local filtering operations. In Section~IV we investigate, how $\EE$ can 
be computed for different types of states. We give explicit formulas for pure 
states, and also show how to compute $\EE$ for 
the family of mixed, Schmidt-correlated 
states. In Section V we show how $\EE$ can be used 
as a lower bound on the concurrence. 
This delivers non-trivial bounds on the concurrence 
for many weakly entangled states.
Finally, we conclude the paper and give technical
 calculations for some of our 
theorems in the Appendix.

\section{Definition of the entanglement parameter}
\label{secdefn}
In this section we introduce a function based on the CMC
that can be used to estimate the amount of entanglement 
in a given quantum state. To start, let us fix the 
notation and introduce the quantities, which we are going
to work with. We consider quantum states  $\vr$ over a 
finite-dimensional bipartite Hilbert space  $\HH_A\otimes\HH_B$, 
where $d_A=\mbox{dim}(\HH_A)$ and $d_B=\mbox{dim}(\HH_B)$ denote 
dimensions of corresponding local spaces. Physical observables 
are described by Hermitian operators. For our purpose we choose 
a complete set of orthogonal observables $A_i$ on $\HH_A$ with
$i=1,...,d_A^2$ and $\tr{A_i A_j}=\delta_{ij}$ and a similar 
set $B_j$ for $\HH_B.$ We will refer to them as local orthogonal 
observables, an example for $d_A=2$ are the (appropriately normalized)
Pauli matrices and the identity. Then, we can consider observables  
on $\HH_A\otimes\HH_B$ defined by
\bear
\{M_{\alpha}\}&=&\{A_i\otimes\eins,\eins\otimes B_j\},\;\;\;\;i=1,\dots, d_A^2,
\nonumber\\ 
&&j=d_A^2+1,\dots,d_A^2+d_B^2,\label{LOOs}
\eear
which then also obey $\tr{M_{\alpha}M_{\beta}} =\delta_{\alpha\beta}$.

The main object of our studies will be covariance matrices (CMs). A
CM of a given bipartite state $\vr$ is defined by the following 
entries
\be
\gamma(\vr)_{\alpha\beta} = 
\halbe\mean{ M_{\alpha} M_{\beta} + M_{\beta} M_{\alpha}}_{\vr} 
- \mean{M_{\alpha}}_{\vr}\mean{M_{\beta}}_{\vr}.
\ee
Choosing the observables as in Eq.~(\ref{LOOs}) one can write the CM 
in a handy block form
\be
\gamma = \begin{pmatrix} A & C\\
                C^T & B
	 \end{pmatrix},
\label{blockcm}
\ee
where $A = \gamma(\vr_A,\{A_i\})$, 
$B = \gamma(\vr_B,\{B_i\})$ are CMs of reduced density 
matrices and $C_{ij}=\mean{A_i\otimes B_j}_{\vr} - \mean{A_i}_{\vr}\mean{B_j}_{\vr}$ 
denote correlations between the two parties.

Before introducing the function that we are going to use for entanglement 
quantification let us state the covariance matrix criterion (CMC). For that, 
recall that a state is separable, if it can be written as a convex combination
of product states, i.e. $\vr=\sum_k p_k \ketbra{a_k}\otimes \ketbra{b_k}$
with some probabilities $p_k$. Then we have:
 
\begin{thm}[Covariance matrix criterion]
\label{CMcrit}
Let $\vr$ be a separable bipartite state.
Then there exist pure states $\ket{\psi_k}\bra{\psi_k}$ in $\HH_A$ 
and $\ket{\phi_k}\bra{\phi_k}$ in $\HH_B$ and convex weights $p_k$ 
such that if we define $\kappa_A=\sum_k p_k \gamma(\ket{\psi_k}\bra{\psi_k})$ 
and $\kappa_B= \sum_k p_k\gamma(\ket{\phi_k}\bra{\phi_k})$ the inequality
\be
\gamma(\vr,\{M_i\}) \geq\kappa_A\oplus\kappa_B
\Longleftrightarrow
\begin{pmatrix} A & C\\
                C^T & B
\end{pmatrix}\geq
\begin{pmatrix} \kappa_A & 0 \\
                0        & \kappa_B
\end{pmatrix}
\label{CMC}
\ee
holds. This means that the difference between left and right 
hand side must be positive-semidefinite. If there are no such 
$\kappa_{A,B}$ then the state $\vr$ must be entangled.
\end{thm}

The proof of this statement can be found in Ref.~\cite{wir}. The 
main task for applying the CMC is  its evaluation, that is, the 
characterization of the matrices $\kappa_A$ and $\kappa_B$. For 
this, several corollaries of the CMC have been derived in 
Refs.~\cite{wir, wirPRA}. As we will use them later, we present 
some of them here, but without any proof. For simplicity, we 
only consider the case $d = d_A = d_B.$

\begin{proposition}[CMC evaluated from traces]
\label{generalcrit}
Let $\vr$ be a state with CM $\gamma$ as in Eq. (\ref{blockcm}).
Then if $\vr$ is separable, we have
\begin{align}
2 \tr{|C|} & 
\leq
\Big(\sum_{i=1}^{d^2}A_{ii}-d+1\Big)+
\Big(\sum_{i=1}^{d^2}B_{ii}-d+1\Big)
\nonumber \\
&=\big[1-\tr{\vr_A^2}\big]
+\big[1-\tr{\vr_B^2}\big],
\label{genercrit}
\end{align}
If this inequality is violated, then $\vr$ must be entangled.
\end{proposition}

\begin{proposition}[CMC and the trace norm of C]
\label{cmctracenorm}
Let $\vr$ be a state with CM $\gamma$ as in Eq.~(\ref{blockcm}).
Then if $\vr$ is separable, we have for the trace norm of $C$
\begin{align}
\Vert C \Vert_{\rm tr}^2 & 
\leq
\big[1-\tr{\vr_A^2}\big]
\big[1-\tr{\vr_B^2}\big],
\label{tracecrit}
\end{align}
If this inequality is violated, then $\vr$ must be entangled.
\end{proposition}

In order to define our entanglement parameter $\EE$, let us reformulate the CMC 
in a slightly different way. Imagine some state $\vr$ is detected as entangled
by the CMC. On the one hand there exist no $\kappa_A$ and $\kappa_B$ as above 
such that $\gamma(\vr)-\kappa_A\oplus\kappa_B\geq 0$. On the other hand 
we can find surely $\kappa^e_A$ and $\kappa^e_B$ and some number $t_e\in[0,1]$ such 
that  $\gamma(\vr)-t_e\kappa^e_A\oplus\kappa^e_B$ is again positive 
semidefinite: 
\be
\gamma(\vr_e)-t_e\kappa^e_A\oplus\kappa^e_B\geq 0.  
\ee 
In the worst case, we can fulfill this inequality by choosing 
$t_e=0.$ For a state that is not detected by the CMC (e.g.~a 
separable state) the parameter $t$ can be chosen to be at least 
one, or even larger than that.

Implementing this idea in Theorem \ref{CMcrit} 
results in an alternative formulation of the CMC:

\begin{thm}[Parameterized CMC]
\label{PCMcrit}
Let $\vr$ be a bipartite state. Assume that we choose pure states 
$\ket{\psi_k}\bra{\psi_k}$ on $\HH_A$ and  $\ket{\phi_k}\bra{\phi_k}$ on $\HH_B$ 
such that 
$\kappa^o_A=\sum_k p_k \gamma(\ket{\psi_k}\bra{\psi_k})$
and
$\kappa^o_B= \sum_k p_k\gamma(\ket{\phi_k}\bra{\phi_k})$ 
are optimal in the sense that
\be
\gamma - t_o\kappa^o_A\oplus\kappa^o_B \geq 0,
\ee
for some $0 \leq t_o\leq 1$, but  
\be
\gamma - t\kappa_A\oplus\kappa_B \ngeq 0,
\mbox{ for all } t > t_o \mbox{ and all } \kappa_A, \kappa_B.
\ee
Then if the state $\vr$ is separable there exist $\kappa^o_A$ and $\kappa^o_B$ 
such that
\be
\max_{t}\{t\leq 1:\gamma - t\kappa^o_A\oplus\kappa^o_B \geq 0\} = 1,
\label{PCMC}
\ee
otherwise the state is entangled.
\end{thm}

This leads to the idea, to use for entangled states the parameter $t_o$ as an 
entanglement parameter. More precisely, we can define:

\begin{defn}[Entanglement parameter]
Let $\vr$ be a bipartite quantum state with CM $\gamma(\vr)$. We 
define a function $V(\vr)$ as
\be
V(\vr)= \max_{t,\kappa_A,\kappa_B}\{t\leq 1 : \gamma(\vr)-
t\kappa_A\oplus\kappa_B\geq 0\}.
\label{covmass}
\ee
The entanglement parameter $\EE(\vr)$ is then defined as
\be
\EE(\vr) = 1 - V(\vr).
\label{entmeas}
\ee
\end{defn}

The parameter $\EE(\vr)$ vanishes for separable states and is larger than zero for 
all states that are detected by the CMC. This function $\EE(\vr)$  is the main 
topic of study in this paper and, as we shall see later, can be used to quantify 
entanglement in quantum states. A similar function has been already used to 
quantify entanglement in infinite dimensional systems, namely Gaussian states 
\cite{giedkecirac}, there this parameter turned out to be an entanglement monotone 
for special operations on special states.

Interestingly, using the parameterized version of the CMC (Theorem \ref{PCMcrit}) and 
Propositions \ref{generalcrit} and  \ref{cmctracenorm} one can immediately give a 
lower bound $\EE(\vr)$. We can formulate:

\begin{proposition}[Bounds on $\EE(\vr)$]
\label{erhobounds}
Assuming that $d=d_A =d_B$ we have in the situation from above  that
\be
\EE(\vr)\geq \frac{\tr{\vr_A^2} + \tr{\vr_B^2} + 2\tr{|C|} - 2}{2 d-2}
\label{ubofe}
\ee
and
\bear
\EE(\vr) &\geq& \frac{1}{d-1}
\Big\{
\frac{\tr{\vr_A^2} + \tr{\vr_B^2}-2}{2} + 
\nonumber
\\
&&+
\sqrt{\tfrac{1}{4}[\tr{\vr_A^2} - \tr{\vr_B^2}]^2+\Vert C \Vert_{\rm tr}^2}
\Big\}.
\label{ubofe2}
\eear
\end{proposition}
{\it Proof.} For the first case, a calculation as in Ref.~\cite{wirPRA} gives 
a parameterized version of Proposition \ref{generalcrit} and results in 
$ 2\tr{|C|}\leq \tr{A+B-t(\kappa_A+\kappa_B)}.$ Using 
$\tr{\gamma(\vr)} = d - \tr{\vr^2}$ (see Ref.~\cite{wirPRA}) gives
\be
t\leq \frac{2 d-\tr{\vr_A^2}-\tr{\vr_B^2}-2\tr{|C|}}{2 d-2}
\label{ubofv}
\ee
and finally Eq.~(\ref{ubofe}). Eq.~(\ref{ubofe2}) can be also directly 
derived from Eq.~(\ref{tracecrit}) and from the calculations in 
Ref.~\cite{wirPRA}.
\qed

\section{Properties of the entanglement parameter $\EE$}
In this section we investigate general properties of the 
function $\EE(\vr)$. Since the function $\EE(\vr)$ should be 
used to quantify entanglement in a given quantum state, two 
of the properties that have to be fulfilled are that 
it is convex and does not change under local unitary 
transformations. Indeed, this is the case:

\begin{lem}[Convexity and invariance under local unitary transformations]
The entanglement parameter $\EE(\vr)$ is invariant under local unitary 
transformations and is convex in the state, that is for
$\vr=p\vr_1+(1-p)\vr_2$
we have that  $\EE(\vr) \leq p \EE(\vr_1) + (1-p) \EE(\vr_2).$
\end{lem}
{\it Proof:} The invariance under local unitary transformations
follows simply from the fact that the CMC is invariant under such
transformations \cite{wir,wirPRA}. In more detail, such transformations
map a set of local orthogonal observables to another set of local 
orthogonal observables, and the CMC does not depend on the choice of 
the observables.

Concerning convexity,  it is sufficient to prove the concavity of 
$V(\vr)$, \emph{i.e.} that for any state $\vr=p\vr_1+(1-p)\vr_2$ 
the  inequality
$
V(\vr)=\tilde{t} \geq p\tilde{t}_1 + (1-p)\tilde{t}_2\equiv t^{\prime}
$
holds, where $\tilde{t}_1 = V(\vr_1)$ and $\tilde{t}_2 = V(\vr_2)$.

To prove this we exploit the connection between the CMC and local uncertainty 
relations (LURs) \cite{lurs}. $V(\vr)=\tilt$ implies that the 
parameterized CMC criterion is fulfilled and there exist $\kappa_A$, 
$\kappa_B$ and $\tilt$ such that
$
\gamma(\vr)-\tilt\kappa_A\oplus\kappa_B\geq 0.
$
According to the Proposition V.2 in Ref.~\cite{wirPRA} this means that if we take 
arbitrary local observables on Alice's and Bob's side ${A}_k\otimes\eins$ 
and $\eins\otimes{B}_k$ such and define positive constants 
$U_A= \min_{\vr}\sum_k\varce{{A}_k}$
and
$U_B= \min_{\vr}\sum_k\varce{{B}_k}$
then
\be
\sum_k\delta^2\left({A}_k\otimes\eins + \eins\otimes
{B}_k\right)_{\vr}\geq \tilde{t}\left(U_A+U_B\right).
\ee
Therefore it suffices to show that $t^{\prime}$ fulfills 
the last inequality as well. Due to the concavity of the variance 
we can write
\bea
\sum_k\delta^2 & \left(A_k\otimes\eins + \eins\otimes B_k\right)_{\vr} 
\geq 
p\sum_k\delta^2\left(A_k\otimes\eins + \eins\otimes
B_k\right)_{\vr_1}
\nonumber
\\
&+ (1-p)\sum_k\delta^2\left(A_k\otimes\eins +
\eins\otimes B_k\right)_{\vr_2}.
\label{lur1}
\eea
Since the states $\vr_1$ and $\vr_2$ both fulfill the CMC 
with the parameters $\tilde{t}_1$ and $\tilde{t}_2$ we can 
write
\bea
p\sum_k\delta^2 & \left(A_k\otimes\eins + \eins\otimes
B_k\right)_{\vr_1} + (1-p)\sum_k\delta^2(A_k\otimes\eins +
\nonumber
\\
+&\eins\otimes B_k)_{\vr_2} \geq \left[\tilde{t}_1p + \tilde{t}_2(1-p) \right]\left(U_A+U_B\right).
\label{lur2}
\eea
Note that $\tilde{t}$ is defined as maximal value of all possible $t$. 
Using (\ref{lur1}) and (\ref{lur2}) this finishes the proof.
\qed

A further important property of entanglement measures is they do not 
increase under local operations assisted with classical communication. 
This condition can be demanded in two different forms 
(see Refs.~\cite{hororeview, plenioreview,loccoa}): 
Minimally, one requires that if $\hat{\vr}$ arises from $\vr$ via some LOCC 
transformation, then $E(\vr) \geq E(\hat{\vr})$ holds. Often, however, 
a stronger condition is required and fulfilled, namely that $E(\vr)$ 
should not increase under LOCC operations \emph{on average}. This means 
that if an LOCC protocol maps $\vr$ onto some states $\vr_i$ with probabilities
$p_i,$ then
\be
E(\vr) \geq \sum_i p_i E(\vr_i),
\label{locconav}
\ee
should hold.

In the following, we will show by an example that $\EE(\vr)$ can 
increase on average under LOCC operations. This does not exclude 
a priori the usability of $\EE(\vr)$ as an entanglement 
monotone (since the  minimal requirement might still hold), however, 
it is a hint that $\EE(\vr)$ might not be an entanglement measure. 
As we will see later, however, $\EE(\vr)$ can be very useful to derive 
lower bounds on the concurrence for mixed states.

\begin{lem}[Increasing one average under LOCC]
There exists a two-qubit state $\vr$ and an LOCC-protocol, 
such that $\EE$ increases on average from zero to a positive
value under this protocol. 
\end{lem}
\emph{Proof.} We prove the statement by providing an explicit example of a 
two-qubit state, which can be found numerically. The idea to find such an example
is as follows: We consider a family of states that was already intensively 
investigated in Refs.~\cite{RudolphCCNC2003pra,wirPRA}. Within this family one can 
find pairs of states $\vr$ and $\vr^{\prime}$ with the same covariance matrix 
but where $\vr$ is entangled, while $\vr^{\prime}$ is not. Hence, $\vr$ cannot 
be detected by the CMC criterion, and $\EE(\vr)$ has to vanish.

It was shown in Refs.~\cite{wir, wirPRA}, however, that after an 
appropriate filtering operation
\be
\vr \mapsto \vr_{\rm filt} = {F_A\otimes F_B \vr F_A^\dagger \otimes F_B^\dagger}
\ee
{\it any} entangled two-qubit state can be detected by the CMC. Hence 
$\EE(\vr_{\rm filt})>0$ and the filtering operation will give rise 
to the desired LOCC operation.

To be more concrete, a numerical example of the aforementioned state 
$\vr$ is
\be
\vr = \left(\begin{array}{cccc}
	0.48508 & 0 & 0 & 0.02094\\
	0 & 0.33 & 0 & 0\\	
	0 & 0 & 0.00067 & 0\\
	0.02094 & 0 & 0 & 0.18425
	\end{array}\right),
\ee
which is not detected by the CMC (see \cite{wirPRA}) but which is clearly 
NPT and hence entangled. The corresponding filter operations are
\bea
F_A &=\blockmatrix{0.16457}{0}{0}{0.98637},\nonumber\\
F_B &=\blockmatrix{0.96526}{0}{0}{0.26128}.
\eea
The final state after $\vr_{\rm filt}$ filtering will be
\bea
\vr_{\rm filt} &=\frac{F_A\otimes F_B \vr^{\prime} F_A\otimes F_B}{Tr\left(F_A\otimes F_B \vr^{\prime} F_A\otimes F_B\right)}\nonumber\\
&= \left(\begin{array}{cccc}
	0.47636 & 0 & 0 & 0.03336\\
	0 & 0.02375 & 0 & 0\\	
	0 & 0 & 0.02364 & 0\\
	0.03336 & 0 & 0 & 0.47626
\end{array}\right)
\eea
and is detected by the CMC, hence $\EE(\vr_{\rm filt})>0$.
Since $\vr$ is not detected, we have $\EE(\vr)=0$.

Using the filter operations $F_A$ and $F_B$ we can now construct 
a POVM type of measurements for Alice and Bob. The complementary 
operations are given by
\bea
F^c_A&=\left(\eins - F_AF_A\right)^{\halbe} = \blockmatrix{0.98637}{0}{0}{0.16457},\nonumber\\
F^c_B&=\left(\eins - F_BF_B\right)^{\halbe} = \blockmatrix{0.26128}{0}{0}{0.96526}.
\eea

With this operations we establish LOCC protocol with four different outcomes
\bea
\vr_1 \equiv \vr_{\rm filt} &\mbox{ with probability } p_1 = 0.02570,\nonumber\\
\vr_2 &\mbox{ with probability } p_2 = 0.17629,\nonumber\\
\vr_3 &\mbox{ with probability } p_3 = 0.46200,\nonumber\\
\vr_4 &\mbox{ with probability } p_4 = 0.33601.
\eea
Important for us is the fact that applying this protocol 
to a state with $\EE(\vr)$ we achieve a state such that 
$\EE(\vr_{\rm filt})$ with 
non-zero probability. Therefore
$
0=\EE(\vr)< \sum_{i=1}^4 p_i \EE(\vr_i),
$
and $\EE(\vr)$ increases on average under LOCC.
\qed

Note that for the provided example one can check the separability of the state 
$\tilde{\vr}=\sum_ip_i \vr_i$ as this state has a positive partial transpose 
and is therefore separable. Consequently, the protocol given is not a 
counterexample to the LOCC condition of the first kind.

\section{Evaluation of $\EE(\vr)$ for pure and 
Schmidt-correlated states}
\label{evaluation}
In this section we compute  $\EE(\vr)$ for pure states and a family 
of mixed states. We start with the case of two-qubits. Then, we 
generalize it to $d$-dimensional systems.

\subsection{Pure states of two qubits}
Using the relations that can be found in Appendix A, it is straightforward 
to calculate the CM of a two-qubit state $\ket{\psi}=\sqrt{\lambda_1}\ket{00}
+\sqrt{\lambda_2}\ket{11}$ with $\lambda_1+\lambda_2=1$. The CM will have the 
familiar block form
\be
\gamma(\ket{\psi})=\left(
\begin{array}{cc}
A & C\\
C^T & B
\end{array}
\right).
\ee
with
\bea
A&=B = \left(
\begin{array}{cccc}
\lambda_1 - \lambda_1^2 & -\lambda_1\lambda_2 & 0 & 0\\
-\lambda_1\lambda_2 &\lambda_2 - \lambda_2^2 & 0 & 0\\
0 & 0 & \halbe & 0\\
0 & 0 & 0 & \halbe
\end{array}
\right), 
\nonumber \\
C &= \left(
\begin{array}{cccc}
\lambda_1 - \lambda_1^2 & -\lambda_1\lambda_2 & 0 & 0\\
-\lambda_1\lambda_2 &\lambda_2 - \lambda_2^2 & 0 & 0\\
0 & 0 & \sqrt{\lambda_1\lambda_2} & 0\\
0 & 0 & 0 & -\sqrt{\lambda_1\lambda_2}
\end{array}
\right).
\eea
The next step in the calculation of the parameter $t$ and therefore of 
the function $\EE(\vr)$ is to find the optimal $\kappa_A\oplus\kappa_B$. 
In the two-qubit case we first guess the correct solution and the prove 
its optimality.

To construct the matrix $\kappa_A\oplus\kappa_B$ we take two product states 
$\ketbra{00}$ and $\ketbra{11}$ and get 
$\kappa_A\oplus\kappa_B=\halbe {\rm diag}\{0,0,1,1,0,0,1,1\}$. 
Then we calculate the $V(\ket{\psi})$ from the condition
$
\gamma - t\kappa_A\oplus\kappa_B \geq 0.
$
This matrix is positive iff
$
1-t\geq 2\sqrt{\lambda_1\lambda_2}
$
and therefore for any
\be 
t\leq 1-2\sqrt{\lambda_1\lambda_2}
\label{schranke}
\ee
we can find $\kappa_A$ and $\kappa_B$ such that 
$\gamma - t\kappa_A\oplus\kappa_B \geq 0$ 
holds.

Note that taking some particular expansion for $\kappa_A\oplus\kappa_B$, 
strictly speaking, does not provide any information about the entanglement, 
except for the case when we are able to find $\kappa_A\oplus\kappa_B$ such 
that $\gamma - t\kappa_A\oplus\kappa_B \geq 0$ for some $t\geq 1$. Then the 
state is not detected by the CMC and $\EE(\vr)=0$. However, we can use the 
Proposition \ref{generalcrit} to prove the following:
\begin{lem}
The upper bound on the parameter $t$ for two qubits provided 
in Eq.~(\ref{schranke}) is tight.
\end{lem}
\emph{Proof: } Directly applying the relation (\ref{ubofv}) to the 
two-qubit case we have $t \leq 1-2\sqrt{\lambda_1\lambda_2}$, 
which coincides with (\ref{schranke}) and therefore gives an 
optimal bound on parameter $t$. Indeed, on the one hand, 
it follows immediately from (\ref{schranke}) that 
if $t\leq 1-2\sqrt{\lambda_1\lambda_2}$ then we can 
find a decomposition $\kappa_A\oplus\kappa_B$ such 
that $\gamma - t\kappa_A\oplus\kappa_B \geq 0$ holds. 
On the other hand, the condition (\ref{ubofv}) implies 
that for all $t$, with $t > 1-2\sqrt{\lambda_1\lambda_2}$ 
and for all 
$\kappa_A$ and $\kappa_B$ the relation 
$\gamma - t\kappa_A\oplus\kappa_B \ngeq 0$ holds.
\qed

According to the last Lemma the function $\EE(\ket{\psi})$ can be calculated exactly for 
two-qubit pure states as
\be
\EE(\ket{\psi}) = 2\sqrt{\lambda_1\lambda_2}.
\label{efunc2qp}
\ee

\subsection{Pure states of two qudits}

To estimate the parameter $t$ for a pure state of two $d$-level systems, 
we follow the same strategy as in the two-qubit case and take the states
$\ketbra{kk}$ for the decomposition of $\kappa_A\oplus\kappa_B$ in order 
to derive the upper bound on the parameter $t$. We make the ansatz
\be
\kappa_A\oplus\kappa_B = \sum_{i=1}^d p_i \gamma(\ket{ii})
\label{optzer}
\ee
with some probabilities $p_i.$

The positive semi-definiteness of the matrix $\gamma -t\kappa_A\oplus\kappa_B$ 
then implies the positive semi-definiteness of $2\times 2$ blocks of the type
\be
X^{ij}_{2\times 2} = \left(\begin{array}{cc}
        \lambda_i+\lambda_j - t(p_i + p_j) & \pm 2\sqrt{\lambda_i\lambda_j}\\
        \pm 2\sqrt{\lambda_i\lambda_j} & \lambda_i+\lambda_j - t(p_i + p_j)
\end{array}\right)
\label{2by2block}
\ee
for all $i<j.$ Therefore, if for all $i < j$ 
\be
t\leq \frac{\left(\sqrt{\lambda_i}-\sqrt{\lambda_j}\right)^2}{p_i+p_j}
\label{tqudits}
\ee
holds, then we can find $\kappa_A$ and $\kappa_B$ such that 
$\gamma - t\kappa_A\oplus\kappa_B \geq 0$ holds.
To achieve the goal and calculate the function $\EE(\ket{\psi})$ we
need to prove that the choice of the expansion of the
$\kappa_A\oplus\kappa_B$ in Eq. (\ref{optzer}) was optimal. 

\begin{lem}[Optimality of the decomposition]\label{lopt}
The optimal expansion for $\kappa_A\oplus\kappa_B$ can always 
be written in a form of the Eq. (\ref{optzer}):
\be
\kappa_A^{opt}\oplus\kappa_B^{opt} = \sum_{i=1}^I p_i \gamma(\ket{ii}).
\label{lemopt}
\ee
\end{lem}
\emph{Proof.}
First, we show that for pure states in Schmidt decomposition 
$\gamma(\ket{\psi})- t\kappa_A\oplus\kappa_B\geq 0$ is equivalent 
to $\gamma(\ket{\psi})- t\kappa\oplus\kappa\geq 0$, 
for some $\kappa$, which can be found explicitly. 
This $\kappa$ can be constructed by choosing 
the product states in a proper way. Indeed, 
note that since the CM of a state in Schmidt 
decomposition is symmetric with respect to 
the interchange of the parties 
($A\leftrightarrow B$) 
the relation $\gamma(\ket{\psi})- t\kappa_B\oplus\kappa_A\geq 0$ 
must hold as well. So let 
$\kappa_A=\sum_{k=1}^{K} p_k \gamma(\ketbra{a_k})$ and  
$\kappa_B=\sum_{k=1}^{K} p_k \gamma(\ketbra{b_k}).$ Then we have
\be
\gamma(\ket{\psi})- 
\frac{t}{2}\left(\kappa_A\oplus\kappa_B + \kappa_B\oplus\kappa_A\right)\geq 0.
\ee
Since 
$\kappa_A\oplus\kappa_B + \kappa_B\oplus\kappa_A 
= \kappa_A\oplus\kappa_A + \kappa_B\oplus\kappa_B$,
the appropriate choice of the product states is
\be
\ket{\eta_k} 
= \left\{\begin{array}{l}
\ket{a_i}\otimes\ket{a_i},\: i=1,\dots,K \;\;(\mbox{for } \kappa_A\oplus\kappa_A), 
\\  
\ket{b_i}\otimes\ket{b_i},\: i=K+1,\dots,2K \;\;(\mbox{for } \kappa_B\oplus\kappa_B).
\end{array}\right.
\ee
Hence we have $\gamma(\ket{\psi})- t \kappa\oplus\kappa\geq 0$, with
\be
\kappa = \sum_{k=1}^{2K} \tilde{p}_k \gamma{\ket{\eta_k}},
\label{preliminaryexp}
\ee
where $\tilde{p}_k = \tfrac{1}{2} p_{(k\; {{\rm mod}}\;K)}$.

Second, because the blocks $D$ in Eq.~(\ref{blockmatrix}) in the Appendix 
are the same, we note that all diagonal elements $D_{ii}$ from $\kappa$ must 
be zero, otherwise only $t=0$ will satisfy $\gamma(\vr)- t\kappa_A\oplus\kappa_B\geq 0$. 
This means that the only states, which can appear in the expansion (\ref{preliminaryexp}) 
are of the form $\ket{\eta_k}=\ket{kk},$ since the $\ket{a_k}$ and $\ket{b_k}$ have to be 
eigenstates of the operators $D_{i}=\ketbra{i}$ (see the Appendix A).
\qed

Having proved the optimality of the expansion of $\kappa_A\oplus\kappa_B$ in 
Eq.~(\ref{optzer}) we can now provide the general formula for the function 
$\EE(\ket{\psi})$ for pure states in the Schmidt decomposition. The value of 
the function $V(\ket{\psi})$ is given by the solution of the following max-min 
problem
\be
\alpha^0 = \max_{\mathcal{P}}\min_{i<j}\frac{\left(\sqrt{\lambda_i}-\sqrt{\lambda_j}\right)^2}{p_i+p_j},\: 1\leq i< j\leq d,
\label{maxminprob1}
\ee
where the first max is taken over all possible probability distributions 
$\PP=\{p_1, p_2, ...\}$. A solution of this problem for the case $d=3$ 
and $d=4$ is given in Appendix and we can summarize:

\begin{proposition}[$\EE$ for pure states]
\label{solutionford34}
(a) If $\ket{\psi}=\sum_{i=1}^{3} \sqrt{\lambda_i} \ket{ii}$
is a pure two-qutrit state, then
\be
\EE(\ket{\psi})=2 \sqrt{\lambda_{i_0} \lambda_{j_0}}+ 
2 \sqrt{\lambda_{i_0} \lambda_{k_0}} - \lambda_{i_0},
\label{qutritpure}
\ee
where $i_0, j_0, k_0$ are pairwise different and 
$j_0, k_0$ are such that 
$(\sqrt{\lambda_{j_0}}-\sqrt{\lambda_{k_0}})^2 \geq (\sqrt{\lambda_{j}}-\sqrt{\lambda_{k}})^2$
for all $j,k.$
\\
(b) If $\ket{\psi}=\sum_{i=1}^{4} \sqrt{\lambda_i} \ket{ii}$
is a pure state in a $4\times4$-system, then 
\be
\EE(\ket{\psi})= \max\{\fe_1,\fe_2,\fe_3 \},
\label{ququadpure}
\ee
where 
\bea
\fe_1 & = 2\sqrt{\lambda_1\lambda_2}+2\sqrt{\lambda_3\lambda_4},
\;\;\;\;
\fe_2 = 2\sqrt{\lambda_1\lambda_3}+2\sqrt{\lambda_2\lambda_4},
\nonumber
\\
\fe_3 & = 2\sqrt{\lambda_1\lambda_4}+2\sqrt{\lambda_2\lambda_3}.
\eea
\end{proposition}
Note that in both cases we have for a maximally entangled state $\EE(\psi)=1.$

\subsection{Schmidt-correlated states}

To conclude the section we consider a family of mixed states, for which 
the introduced function $\EE(\vr)$ can be also computed exactly. 
These states are called {Schmidt-correlated (SC) states} in 
the literature \cite{scstates}. By definition, SC states are a
mixture of states that share the same Schmidt basis
\bea
\vr_{SC} &= \sum_{u=1}^N q_u\ket{\psi_u}\bra{\psi_u},\mbox{ with}\\
\ket{\psi_u} &= \sum_{i=1}^{d} \sqrt{\lambda_i^{(u)}}\ket{ii}.
\eea
SC states can be written in computational basis directly as
\be
\vr_{SC} = \sum_{ij}\vr_{ij}\ket{ii}\bra{jj}, \mbox{ with } \vr_{ij} = \sum_u q_u\sqrt{\lambda^{(u)}_i\lambda^{(u)}_j}.
\ee 

As in the case of pure states, we find for the SC states the optimal 
decomposition of $\kappa_A\oplus\kappa_B:$

\begin{lem}[Optimality for SC states]\label{sclopt}
In the case of SC states the optimal decomposition of 
$\kappa_A\oplus\kappa_B$ for the estimation of the 
parameter $\EE(\vr)$ can always be written in the 
form of Eq.~(\ref{optzer}):
\be
\kappa_A^{opt}\oplus\kappa_B^{opt} = \sum_{i=1}^d p_i \gamma(\ket{ii})
\label{lemoptsc}.
\ee
\end{lem}
\emph{Proof: }
There were two essential ingredients in the proof of the Lemma \ref{lopt}. 
First, we used the fact that the CM of a state, written in Schmidt decomposition, 
is invariant under interchange of parties. Obviously the same invariance does 
also hold for SC states. Second, we used the fact, that all blocks $D$ of the 
CM are the same. Using the formulae of the Appendix A one easily verifies that $D^{A}_{SC}=D^{B}_{SC}=D^{C}_{SC}$.
\qed

For these states the problem of calculating the function $\EE(\vr)$ reduces to 
the max-min problem in Eq.~(\ref{maxminprob1}). This is due to the fact that diagonal 
elements of the covariance matrix have a pretty simple form for $\vr_{SC}$. 
Indeed, using the formulae from the Appendix A we calculate directly:
\bea
&(D^{A/B/C}_{SC})_{ij} 
 = \vr_{ii} \delta_{ij} - \vr_{ii}\vr_{jj},\;\; 1\leq i\leq d,
\nonumber\\
&X^{A/B}_{SC}  = Y^{A/B}_{SC} = \halbe\mbox{diag}\{\vr_{ii}+\vr_{kk}\},
\;\;1\leq i<k\leq d,
\nonumber\\
&X^{C}_{SC} =-Y^{C}_{SC}= \mbox{diag}\{\vr_{ik}\},\: 1\leq i<k\leq d.
\eea
The $2\times 2$ blocks in Eq.~(\ref{2by2block}) will then take 
the form
\be
B^{ij}_{2\times 2} = \left(\begin{array}{cc}
        \vr_{ii}+\vr_{jj} - t(p_i + p_j) & \pm 2\vr_{ij}\\
        \pm 2\vr_{ij} & \vr_{ii}+\vr_{jj} - t(p_i + p_j)
\end{array}\right),
\label{2by2blockSC}
\ee
which leads to the following max-min problem for $V(\vr_{SC})$
\bea
V(\vr_{SC}) &= \max_{\mathcal{P}}\min_{i<j}\frac{\vr_{ii}+\vr_{jj}-2\vr_{ij}}{p_i+p_j}\nonumber\\ 
&= \max_{\mathcal{P}}\min_{i<j}\frac{\sum_k q_k\left( \sqrt{\lambda^{(k)}_i} - \sqrt{\lambda^{(k)}_j} \right)^2}{p_i+p_j}.
\eea
This problem can be effectively solved numerically or with the methods 
of the Appendix B and its solution gives the exact value of the function 
$\EE(\vr_{SC})$.
For two qubits one finds 
\be
\EE(\vr_{SC})=2\sum_k q_k\sqrt{\lambda^{(k)}_0\lambda^{(k)}_1}
\ee
as a nice analytical expression.
%

\section{The entanglement parameter $\EE(\vr)$ as a lower bound on 
the concurrence}

In this section we demonstrate that the function $\EE(\vr)$ can be used 
to estimate the amount of entanglement in a quantum state. More specifically, 
we show how it delivers a lower bound on the concurrence, which is a well 
known measure of bipartite entanglement. For bipartite pure states in a 
$d\times d$-system the concurrence is defined as \cite{conc1,conc2,conc3}:
\be
C(\ket{\psi})=\sqrt{\frac{d}{d-1}}\sqrt{1-\tr{\vr_A^2}}.
\ee
In this definition, we introduced already a prefactor which guarantees 
that $0 \leq C \leq 1,$  this will turn out to be useful for our purposes.

The concurrence is then extended to mixed states by the convex-roof 
construction
\be
C(\vr) = \min_{p_i,\ket{\psi_i}}\sum_i p_iC(\ket{\psi_i}),
\ee
where the minimization is taken over all possible decompositions of the state 
$\vr=\sum_ip_i\ketbra{\psi_i}$. Of course, this minimization is quite difficult 
to perform, and only for two-qubits a complete solution is known \cite{conc2}. Therefore, 
it is desirable to have at least some lower bounds on the concurrence.

The idea of obtaining lower bounds on $C$ from $\EE$ is as follows: Let us assume that 
one can prove a lower bound like
\be
C(\ket{\psi}) \geq \alpha \EE(\ket{\psi}) + \beta
\label{chentrick}
\ee
for pure states only with some constants $\alpha,\beta$ and $\alpha \geq 0.$
Then, since $\EE$  is convex, the right hand side of 
Eq.~(\ref{chentrick}) is convex, too. By definition, the convex roof is the 
largest convex function which coincides with $C$ on the pure states. Consequently, 
$C(\vr) \geq \alpha \EE(\vr) + \beta$ holds for all mixed states, too. This 
trick has already been employed in several works to obtain lower bounds on entanglement 
measures \cite{chenbounds,otherconvexbounds}. However, as the CMC 
detects many bound entangled states where other 
criteria fail \cite{wirPRA}, 
our results will deliver entanglement estimates for states, where the 
other methods fail.

\subsection{Two qubits}

Using the Schmidt decomposition, one can express the concurrence 
for pure states in terms of Schmidt coefficients as
\be
C(\ket{\psi}) = \sqrt{\frac{2d}{d-1}} \sqrt{\sum_{i<j} \lambda_i\lambda_j}.
\label{wc}
\ee
Comparing Eq.~(\ref{wc}) and Eq.~(\ref{efunc2qp}) from the Section \ref{evaluation} 
we see that the concurrence and the function $E(\ket{\psi})$ coincide for 
on two-qubit pure states
\be
\EE(\ket{\psi}) = 2\sqrt{\lambda_1\lambda_2} = C(\ket{\psi}).
\ee
Consequently, $C(\vr)\geq \EE(\vr)$ holds for any mixed state. Note, however, 
that for the special case of two qubits one can calculate the concurrence also 
directly for mixed states \cite{conc2}. 

\subsection{Two qutrits}

Using the solution of the problem (\ref{maxminprob1}) it is possible to derive 
a lower bound on concurrence for pure states of two $d$-level systems.  Before we
proceed, note that \cite{chenbounds}
\be
C(\ket{\psi})=\sqrt{\frac{2d}{d-1}} \sqrt{\sum_{i<j} \lambda_i\lambda_j}
\geq \frac{2}{d-1}\sum_{i<j}\sqrt{\lambda_i\lambda_j},
\label{nottight}
\ee
This follows from the fact that
\bea
&\sum_{i<j}\lambda_i\lambda_j
=\frac{1}{d(d-1)}\sum_{i<j}\sum_{k<l}(\lambda_i\lambda_j+\lambda_k\lambda_l)
\label{concestim}
\\
&\geq
\frac{2}{d(d-1)} \sum_{i<j}\sum_{k<l}\sqrt{\lambda_i\lambda_j\lambda_k\lambda_l}
=\frac{2}{d(d-1)}\Big[\sum_{i<j}\sqrt{\lambda_i\lambda_j}\Big]^2.
\nonumber
\eea
For two qutrits $\EE(\ket{\psi})$ is given by Eq.~(\ref{qutritpure}).
We have that
\bea
2&\sqrt{\lambda_i \lambda_j}
+2\sqrt{\lambda_i \lambda_k}
-\lambda_i
= 
2\sqrt{\lambda_i \lambda_j}
+2\sqrt{\lambda_i \lambda_k}
+2\sqrt{\lambda_j \lambda_k}
\nonumber
\\ 
&-2\sqrt{\lambda_j \lambda_k}
-1+\lambda_j +\lambda_k
\nonumber
\\
&
\leq 2 C(\ket{\psi}) + (\sqrt{\lambda_j}-\sqrt{\lambda_k})^2 -1
\nonumber
\\
&
\leq 2 C(\ket{\psi}).
\eea
Hence we have for mixed two-qutrit states
\be
C(\vr) \geq \frac{\EE(\vr)}{2}.
\ee
Using the results from Proposition \ref{erhobounds} we have, for instance,
\bear
C(\vr) &\geq& \frac{1}{4}
\Big\{
\frac{\tr{\vr_A^2} + \tr{\vr_B^2}-2}{2} + 
\nonumber
\\
&&+
\sqrt{\tfrac{1}{4}[\tr{\vr_A^2} - \tr{\vr_B^2}]^2+\Vert C \Vert_{\rm tr}^2}
\Big\},
\label{cbound33}
\eear
which is an easily computable lower bound that delivers non-trivial estimates 
for many weakly entangled states.
\subsection{$4\times 4$ systems}
In this case $\EE(\ket{\psi})$ is given by Eq.~(\ref{ququadpure}).
We can directly estimate:
\bea
\EE(\ket{\psi}) & = \max\{\fe_1,\fe_2,\fe_3 \}
\leq 2\sqrt{\lambda_1\lambda_3}+2\sqrt{\lambda_2\lambda_4}\\ 
&+ 2\sqrt{\lambda_1\lambda_2}+2\sqrt{\lambda_3\lambda_4} + 2\sqrt{\lambda_1\lambda_4}+2\sqrt{\lambda_2\lambda_3}\nonumber\\ 
&\leq 3C(\ket{\psi})
\label{lower}
\nonumber
\eea
and hence for arbitrary mixed states
\be
C(\vr)\geq \frac{1}{3}\EE(\vr).
\ee

\subsection{Examples}

Let us discuss the strength of these lower bounds by considering 
some examples. Let us first consider Bell-diagonal two-qubit states. 
For them, the reduced states $\vr_A$ and $\vr_B$ are maximally 
mixed, and then Proposition \ref{erhobounds} delivers the bound 
$C(\vr) \geq \tr{|C|}-1/2.$ On the other hand, it is known that
for Bell diagonal states the concurrence is given by 
$C(\vr)=2\lambda_{\rm max}-1$, where $\lambda_{\rm max}$ is the 
maximal eigenvalue, i.e., the maximal overlap with some Bell state 
\cite{conc1}. Noting that $\lambda_{\rm max}= [1+2\tr{|C|}]/4$ (this can be 
easily seen if the closest Bell state is the singlet state and we
take appropriately normalized Pauli matrices as observables in the definition
of the matrix $C$), one finds that our lower bound is tight for 
Bell diagonal states.

For general two-qubit states, the lower bound cannot be tight, as 
they are entangled two-qubit states, which are not detected by the 
CMC. On the other hand, any full rank two qubit state can be brought
to a Bell-diagonal state by filtering operations. Since it is known
how the concurrence changes under filtering operations \cite{frank}, one could
use the filtering and our lower bound to determine the concurrence
for arbitrary two-qubit states.

For two qutrits, our bound is not tight for states like
$\ket{\psi}=(\ket{00}+\ket{11}+\ket{22})/\sqrt{3}$
or $\ket{\psi}=(\ket{00}+\ket{22})/\sqrt{2},$ however, for 
the latter the reason lies in the fact that the bound 
(\ref{nottight}) is not tight.
On the other hand, the presented method delivers 
nontrivial lower bounds for many bound entangled states 
(such as the the family of chessboard states), as many 
states of this type are detected by the CMC \cite{wirPRA}, but 
not by the PPT or CCNR criterion (which means that the methods 
from Ref.~\cite{chenbounds} must fail). Similarly, our methods can be 
used to estimate the entanglement of bound entangled states 
for $4 \times 4$-systems.

\section{Conclusion}
In conclusion, we have introduced an entanglement parameter $\EE$ 
that quantifies the violation of the covariance matrix criterion. 
We have shown that this parameter is convex and invariant under 
local rotations, but it can increase on average under local operations 
and classical communication. Most 
importantly the parameter $\EE$ can be used to deliver lower bounds
on the concurrence.

For future work, it would be interesting to connect $\EE$ 
to other entanglement measures, such as the entanglement of 
formation \cite{plenioreview}. Even more interesting, would 
be an extension of the covariance matrix criterion to the 
multipartite case and a definition of a similar entanglement parameter there. This 
could help to quantify entanglement in multipartite systems, 
where much less is known compared to bipartite systems.

We thank Jens Eisert, Bastian Jungnitsch, Matthias Kleinmann, 
and S\"onke Niekamp for discussions. Especially we thank Philipp
Hyllus for discussions and comments on the manuscript. This work 
has been supported by the FWF (START Prize) and the 
EU (OLAQUI, QICS, SCALA).

\setcounter{thm}{0}
\setcounter{equation}{0}
\setcounter{figure}{0}
\renewcommand{\theequation}{A-\arabic{equation}}
\renewcommand{\thefigure}{A-\arabic{figure}}
\renewcommand{\thethm}{A-\arabic{thm}}

\section*{{\bf APPENDIX A.}}\label{appena}
Here we calculate symmetric block CM of a pure bipartite state, 
which is written in the Schmidt decomposition 
$\ket{\psi}=\sum_i\sqrt{\lambda_i}\ket{i_A}\otimes\ket{i_B}$. 
Consider $d_A=d_B=d$. As it is proven in \cite{wirPRA} we can 
choose the basis in the operator spaces $\BB(\HH_A)$ and $\BB(\HH_B)$ 
arbitrarily for applying the CMC. In this case it is convenient to choose 
the local orthogonal observables
\bea
D_i &= \ketbra{i}, \;\;\; i = 1, \dots ,d, \\
X_{i,j} &= \frac{1}{\sqrt{2}} (\ket{i}\bra{j}+\ket{j}\bra{i}), \;\;\; 1
\leq  i < j \leq d, \\
Y_{k,l} &= \frac{i}{\sqrt{2}} (\ket{k}\bra{l}-\ket{l}\bra{k}), \;\;\; 1
\leq  k<l \leq d,
\label{standbas}
\eea
which satisfy following anticommutation relations:
\bea
\Akommut{D_i}{D_{j}}&= \delta_{ij}\left(\ket{i}\bra{j} + \ket{j}\bra{i}\right),
\;\;\; \Akommut{D_i}{X_{ij}}= X_{ij} ,
\nonumber\\
\Akommut{D_i}{Y_{ij}}&= Y_{ij} ,
\;\;\;\;\;\;\;\;\;\;\;\;\;\;\;\;\;\;\;\;\;\;\;\;\;\; \Akommut{X_{ij}}{Y_{ij}}= 0  ,
\nonumber\\
\Akommut{X_{ij}}{X_{ij}}&= D_i + D_j  ,
\;\;\;\;\;\;\;\;\;\;\;\;\;\;\;\;\;\; \Akommut{Y_{ij}}{Y_{ij}}= D_i + D_j.
\label{akommrels}
\eea
Note that this is not the complete set of relations, however other 
relations will not give any contribution to the CM and hence we leave 
them out here.

The mean values for the state $\ket{\psi}$ are given by
\bea
&\mean{X^A_{ij}\otimes\eins}=\mean{\eins\otimes X^B_{ij}}=
\mean{Y^A_{ij}\otimes\eins}=\mean{\eins\otimes Y^B_{ij}}=0\nonumber\\
&\mean{D^A_{i}\otimes\eins}=\mean{\eins\otimes D^B_{i}}=\lambda_i\nonumber\\
&\mean{\Akommut{D^A_i}{D^A_{j}}\otimes\eins} = (\lambda_i+\lambda_j)\delta_{ij}
\label{meanvals}
\eea
The blocks $A,B$ and $C$ of the $\gamma^S(\ket{\psi})$ can be therefore 
written as $3\times 3$ block matrices. Because of the relations 
(\ref{akommrels}) and (\ref{meanvals}) a lot of terms in these 
blocks will be equal to zero an we have the structure
\be
A,B,C = \left(
\begin{array}{ccc}
D^{A/B/C} & 0 & 0\\
 0 & X^{A/B/C} & 0\\
0 & 0 & Y^{A/B/C}
\end{array}
\right),
\label{blockform}
\ee
Since the off-diagonal terms can be calculated straightforward
\bea
&\mean{D_i^A\otimes D_j^B} - \mean{D_i^A}\mean{D_j^B}= \lambda_i\delta_{ij} - \lambda_i\lambda_j,\nonumber\\
&\mean{D_i^A\otimes X_{qr}^B}=
\mean{D_i^A\otimes Y_{qr}^B}=
\mean{X_{pq}^A\otimes Y_{rs}^B}=0,\nonumber\\
&\mean{X_{pq}^A\otimes X_{rs}^B}= \sqrt{\lambda_p\lambda_q}\delta_{pr}\delta_{qs},\nonumber\\
&\mean{Y_{pq}^A\otimes Y_{rs}^B}= -\sqrt{\lambda_p\lambda_q}\delta_{pr}\delta_{qs},
\eea
we can write the blocks in (\ref{blockform})
as follows
\bea
D=D^{A/B/C}_{ij} &= \lambda_i\delta_{ij} - \lambda_i\lambda_j, \nonumber\\
X= X^{A/B} &= \halbe\mbox{diag}\{\lambda_i+\lambda_k\}, 1\leq i<k\leq d, \nonumber\\
Y= Y^{A/B} &= \halbe\mbox{diag}\{\lambda_i+\lambda_k\}, 1\leq i<k\leq d,\nonumber\\
X^C &= \mbox{diag}\{\sqrt{\lambda_p\lambda_q}\}, 1\leq p<q\leq d,\nonumber\\
Y^C &= \mbox{diag}\{-\sqrt{\lambda_p\lambda_q}\}, 1\leq p<q\leq d.\label{mateqns}
\eea
Finally, we arrive at the general form of the CM for a pure state as a function its Schmidt coefficients:
\be
\gamma^S(\ket{\psi}) = \left(\begin{array}{cccccc} D & 0 & 0 & D & 0 & 0\\
                           0 & X & 0 & 0 & X^C & 0\\
                           0 & 0 & Y & 0 & 0 & Y^C\\
                           D & 0 & 0 & D & 0 & 0\\
                           0 & X^C & 0 & 0 & X & 0\\
                           0 & 0 & Y^C & 0 & 0 & Y\\
               \end{array}\right)
\label{blockmatrix}
\ee
with the blocks given in Eq.~(\ref{mateqns}).
\setcounter{thm}{0}
\setcounter{figure}{0}
\setcounter{subfigure}{0}
\setcounter{equation}{0}
\renewcommand{\thefigure}{B-\arabic{figure}}
\renewcommand{\theequation}{B-\arabic{equation}}
\renewcommand{\thethm}{B-\arabic{thm}}
\section*{{\bf APPENDIX B.}}\label{appenc}

\begin{figure}[t!]
\subfigure{
\includegraphics[width=0.45\columnwidth]{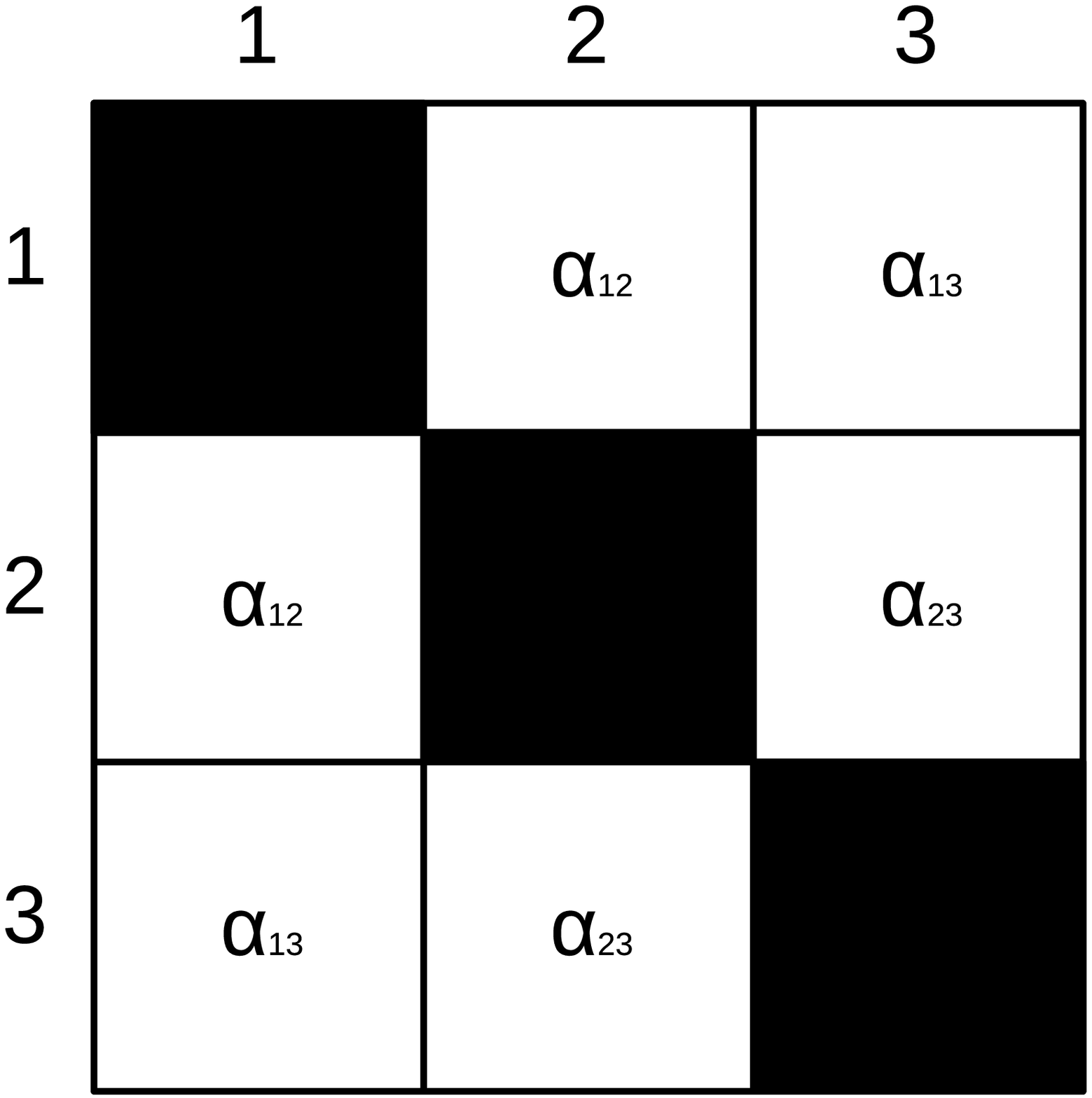}
}
\subfigure{
\includegraphics[width=0.45\columnwidth]{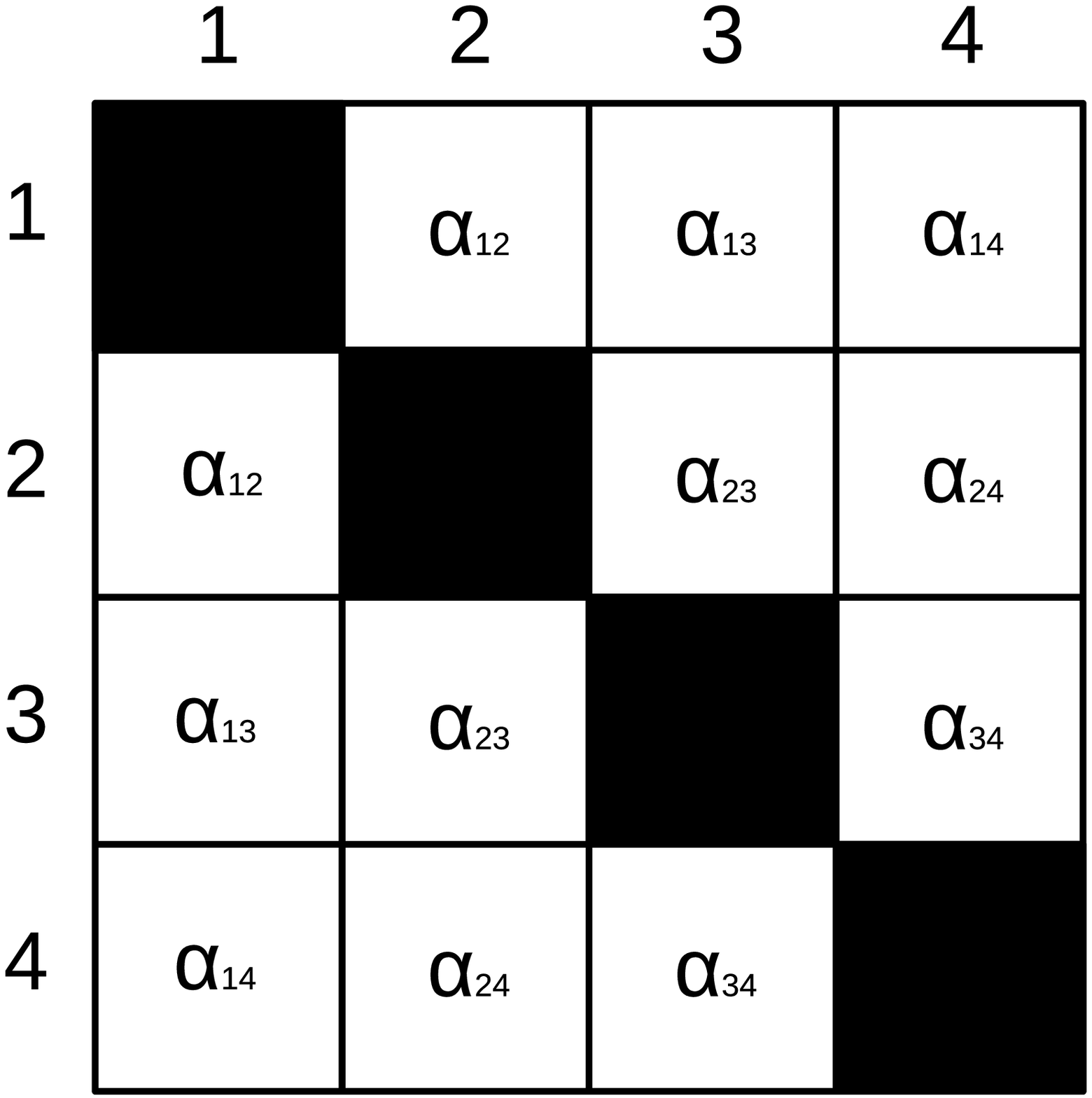}
}
\caption{(a) In the case $d=3$ there are only three elements
$\alpha_{ij}$. These elements are written in the form of tableaux 
in order to visualize the problem considered. (b) Tableaux of 
the elements $\alpha_{ij}$ in the max-min problem for $d=4$.}
\label{tableaux}
\end{figure}

In this Appendix we discuss the possible ways of solving the 
max-min problem:
\be
\tilt = \max_{\mathcal{P}}\min_{i<j}
\frac{\left(\sqrt{\lambda_i}-\sqrt{\lambda_j}\right)^2}{p_i+p_j},\: 1
\leq i<j\leq d.
\label{maxminprob}
\ee
We consider the cases $d=3$ and $d=4$. We define
\bea
b_{ij}&\equiv \left(\sqrt{\lambda_i}-\sqrt{\lambda_j}\right)^2,
\label{bijdef}
\\
\alpha_{ij}&\equiv \frac{b_{ij}}{p_i+p_j} = \alpha_{ji}.
\eea
For $d=3$ there are only three different $\alpha$'s that can 
be arranged in a tableaux as in Fig.~\ref{tableaux}(a).

The properties of the solution can be summarized as follows:

\begin{lem}\label{lemd3}
(a) Consider the optimization problem in Eq.~(\ref{maxminprob}) 
for d=3 with the only assumption that $b_{ij}\geq 0.$ 
Let $j_0,k_0$ be such that
\be
b_{j_0 k_0} = \max\{b_{jk}\}
\label{bindex}
\ee
Then the optimal solution $\alpha^0$ is given by 
\be
\alpha^0 = \min\left\{\alpha^I,\alpha^{II}\right\},
\ee
where 
\bea
\alpha^I &= \frac{1}{2}(b_{12}+b_{13}+b_{23})
\\
\alpha^{II} &= b_{i_0,j_0} +b_{i_0,k_0}
\label{lemd3sol}
\eea
with $j_0 \neq i_0 \neq k_0.$ 
\\
(b) For the same problem, if the $b_{ij}$ are given via 
Eq.~(\ref{bijdef}) as functions of Schmidt coefficients
and fulfill therefore further restrictions, the optimum 
is always given by 
\be
\alpha^0 = \alpha^{II}
=
1 + \lambda_{i_0} 
- 2\sqrt{\lambda_{i_0}\lambda_{j_0}} - 2\sqrt{\lambda_{i_0}\lambda_{k_0}}
\ee
Then we also have that 
$\alpha^{II}=\min_{ijk}\{1 + \lambda_{i} - 2\sqrt{\lambda_{i}\lambda_{j}} - 2
\sqrt{\lambda_{i}\lambda_{k}} \}$ where the $i,j,k$ are pairwise different.
\end{lem}
\emph{Proof: } 
(a) Let us first assume only that $b_{ij}\geq 0$.
In the max-min problem (\ref{maxminprob}) the maximization is taken over 
all possible probability distributions. It is convenient to distinguish 
two cases:

\emph{Case 1:} 
The optimal probability distribution does not have any zero elements. 
Assume $\mathcal{P}_0$ is the optimal distribution and $p_i^0\neq 0$ 
$\forall i$. We often drop the index $^0$ in the following for simplicity.
We show that this optimal distribution necessarily has to be such that 
$\alpha_{12}=\alpha_{13}=\alpha_{23}=\alpha^0$, otherwise the 
optimality is violated. Indeed, assume that this is not case. Then, 
without loss of generality, we can write $\alpha_{12}\leq\alpha_{13}\leq
\alpha_{23}$, where one of the inequalities must be strict. Now
consider some distribution $\mathcal{P}^{\prime}$ such that 
\bea
p^{\prime}_1 &= p_1 - 2\varepsilon, \;\;\;
p^{\prime}_2 = p_2 +\varepsilon, \;\;\;
p^{\prime}_3 = p_3 +\varepsilon,
\eea 
with some $\varepsilon >0$. 
The coefficients $\alpha_{ij}$ will change and
become according to the new distribution $\mathcal{P}^{\prime}$
\be
\alpha^{\prime}_{12} > \alpha_{12},\;\;\;
\alpha^{\prime}_{13} > \alpha_{13},\;\;\;
\alpha^{\prime}_{23} < \alpha_{23}.
\label{probchange}
\ee 
Since the parameter $\varepsilon$ can be chosen arbitrarily small 
the number $\alpha^{\prime}_{12}$ will be still the minimal one, 
\emph{i.e.} $\alpha^{\prime}_{12}=\min_{i<j}\alpha^{\prime}_{ij}$. 
But $\alpha^{\prime}_{12} > \alpha_{12}$. Consequently the 
distribution $\mathcal{P}^{\prime}$ gives a bigger minimum of 
the set $\{\alpha_{ij}\}$ than the distribution $\mathcal{P}_0$, 
which contradicts the assumption that $\mathcal{P}_0$ is optimal. Hence we
conclude that $\alpha_{12}=\alpha_{23}$ must hold, which implies 
$\alpha_{12}=\alpha_{13}=\alpha_{23}.$

Having established that if $\mathcal{P}_0$ is optimal and contains no
zero elements, then $\alpha_{12}=\alpha_{13}=\alpha_{23}=\alpha^0
$ holds, we can calculate $\alpha^0$ explicitly. We have
\bea
\alpha^0 &= \frac{b_{12}}{p_1+p_2} 
= \frac{b_{13}}{p_1+p_3} = \frac{b_{23}}{p_2+p_3}.
\eea
Multiplying by the denominators summing up these equations gives
\be
2\alpha^0(p_1+p_2+p_3) = b_{12} + b_{13} + b_{23}.
\ee
Because $p_1+p_2+p_3=1$ we arrive at
\be
\alpha^I \equiv \alpha^0= \frac{1}{2}(b_{12}+b_{13}+b_{23}).
\label{solcase1}
\ee

\emph{Case 2:} 
The optimal probability distribution $\mathcal{P}_0$ 
has at least one zero element. This means that one 
$\alpha_{ij}=b_{ij}$ independently of the two free parameters
of the probability distribution (since $p_i+p_j=1$). 
We can distinguish three 
cases, and assume for definiteness $b_{12} \leq b_{13}\leq b_{23}.$

(i) If $\alpha_{12}=b_{12}$ (that is, $p_3=0$),
then clearly $\alpha_{ij}\geq b_{ij}$ 
for $i,j=1,3$ and $i,j=2,3.$ Then we have $\min\{\alpha_{ij}\}=\alpha_{12}.$
But then decreasing one of the $p_{1}$ or $p_2$ and increasing consequently
$p_3$ will lead to an increasing of $\alpha_{12}$ and a better solution which
belongs to case 1. So a solution with $\alpha_{12}=b_{12}$ can never be optimal.

(ii) If $\alpha_{13}=b_{13}$ the optimal probability distribution 
has to be such that $\alpha_{13}  \leq \alpha_{12}$ and 
$\alpha_{13} \leq \alpha_{23}.$ But as in the case (i) one can 
directly see that this leads to case 1 and can never be optimal.

(iii) Finally, consider the case $\alpha_{23}=b_{23}.$ Then, one can see 
as in case 1 one can achieve  $\alpha_{12}=\alpha_{13}$ without giving up
optimality. More precisely, the optimal probability distribution
has to fulfill this from the beginning (if $\alpha_{12}$ and $\alpha_{13}$ 
are the minima) or it can be achieved (if $\alpha_{23}$ is the minimum).

This leads as in case 1 to the conclusion, that we have
\bea
\alpha_{12} &= \frac{b_{12}}{p_2} = \frac{b_{13}}{p_3}
\Rightarrow
\alpha_{12} = b_{12} + b_{13},
\eea
and consequently
\be
\alpha^{II} \equiv \alpha_{12} = b_{12} + b_{13} = 1 + \lambda_1 - 2\sqrt{\lambda_1\lambda_2}
- 2\sqrt{\lambda_1\lambda_3}.
\label{solal2}
\ee
However, it is not yet clear what the $\min\{\alpha_{ij}\}$ is. 
Two cases can be distinguished:

(iiia)
If $\alpha_{12} \geq \alpha_{23}=b_{23}$ one would 
take $\min\{\alpha_{ij}\} =\alpha_{23}=b_{23},$  
but then, one can improve it further as in the cases 
(i) and (ii) by going to the case I and taking finally 
$\alpha^{I}$ from Eq.~(\ref{solcase1}). Note that
$\alpha^I = (\alpha^{II}+b_{23})/2.$
Therefore, if $\alpha^{II}= \alpha_{12} \geq \alpha_{23}=b_{23}$ 
one has also that $\alpha^{I}\leq \alpha^{II},$ so 
effectively one takes $\min\{\alpha^{I},\alpha^{II}\}.$

(iiib)
If $\alpha_{12} < \alpha_{23}=b_{23}$ we take $\min\{\alpha_{ij}\} = \alpha^{II}$ 
and going to case 1 does not help. But in this case, we have 
$\alpha^{I}\geq \alpha^{II},$ so effectively one takes again
$\min\{\alpha^{I},\alpha^{II}\}.$

Finally, let us discuss shortly the meaning of the 
choice $j_0$ and $k_0$ in Eq.~(\ref{bindex}) as one 
may consider also $\alpha^{II}_{j,k}$ in Eq.~(\ref{lemd3sol}) 
with other indices. However, one can directly compute that
$\alpha^{II}_{i,j} < \alpha^I$ is equivalent to 
$b_{ij}+b_{ik}< b_{jk}$ and this can only be true, if $j$ and
$k$ are chosen as in Eq.~(\ref{bindex}). In other words, 
the $\alpha^{II}_{j,k}$ for other indices than $j_0,k_0$
can never contribute and one could alternatively write that
$\alpha_0=\min\{\alpha^I,\alpha^{II}_{12},\alpha^{II}_{13}, \alpha^{II}_{23}\}.$

(b) Let us now assume that the $b_{ij}$ stem from Schmidt coefficients
as in Eq.~(\ref{bijdef}). We know from the previous discussion that 
we have to take $\alpha^{II}$ iff $b_{i_0 j_0}+b_{i_0 k_0} \leq  b_{j_0 k_0}.$
In terms of the Schmidt coefficients, this  implies that
\be 
(\sqrt{\lambda_{j_0}}-\sqrt{\lambda_{k_0}})^2 
\geq (\sqrt{\lambda_{i_0}}-\sqrt{\lambda_{j_0}})^2+(\sqrt{\lambda_{i_0}}-\sqrt{\lambda_{k_0}})^2.
\ee
This, however, is true for any triple of positive real numbers $\sqrt{\lambda_{\nu}},$ if $j_0$ and $k_0$
are chosen as in Eq.~(\ref{bindex}). Then, its also clear that the $\alpha^{II}$ chosen
is minimal among all the $b_{ij}+b_{ik}.$
\qed

Further, we discuss the case $d=4$. The elements $\alpha_{ij}$ are again
embedded in a tableaux as in Fig.~\ref{tableaux}(b). We begin with studying of
properties of the optimal probability distribution $\mathcal{P}_0$.
Suppose as in the case $d=3$ that $\alpha_{ij}^0$ correspond to the
optimal probability distribution $\mathcal{P}_0$ and that
$\alpha_{12}^0=\min_{ij}\{\alpha_{ij}^0\}$. We can formulate:

\begin{lem}
\label{lemd4}
The solution of the max-min problem (\ref{maxminprob1}) 
for $d=4$ is given by 
\be
\alpha^0=\min\{\fa^{I},\fa^{II},\fa^{III}\},
\ee
where
\bea
\fa^I=1 - 2\sqrt{\lambda_1\lambda_2}-2\sqrt{\lambda_3\lambda_4},
\nonumber
\\
\fa^{II}=1 - 2\sqrt{\lambda_1\lambda_3}-2\sqrt{\lambda_2\lambda_4},
\nonumber
\\
\fa^{III}=1 - 2\sqrt{\lambda_1\lambda_4}-2\sqrt{\lambda_2\lambda_3}.
\label{lem4deq}
\eea
\end{lem}

\emph{Proof:}
The proof proceeds in several steps.
\\
{\it Step 1.} Let us first consider optimal probability distributions 
$\PP_0= \{p_1,p_2,p_3,p_4\}$ where all $p_i$ are nonzero. 
In this case we show that for $\alpha^0=\min\{\alpha_{ij}\}$ 
at least one of the three equations must hold:
\bea
\alpha^0&= \alpha_{12}=\alpha_{34},\nonumber \\
\alpha^0&= \alpha_{13}=\alpha_{24} \nonumber \\
\alpha^0&= \alpha_{14}=\alpha_{23}.
\label{threecandidates}
\eea

The idea of the proof is similar to the proof of Lemma \ref{lemd3}: 
we consider small perturbations of the optimal probability distribution 
$\PP_0$ that increase the minimal element $\alpha^0$ and therefore 
destroy the optimality if some additional constraints are not fulfilled. 
As we will see, these constraints will give us the conditions Eq.~(\ref{threecandidates}).

Let us assume for definiteness that the 
optimal $\alpha^0$ is given by $\alpha_{12}$.
We can consider the following four transformations
of the $p_i:$
\bear
T_1: && 
p_1'=p_1-3\varepsilon, \;\;\; p_i'= p_i+\varepsilon \mbox{ for } i \neq 1,
\nonumber
\\
T_2: && 
p_2'=p_2-3\varepsilon, \;\;\; p_i'= p_i+\varepsilon \mbox{ for } i \neq 2,
\nonumber
\\
T_3: && 
p_3'=p_3+3\varepsilon, \;\;\; p_i'= p_i-\varepsilon \mbox{ for } i \neq 3,
\nonumber
\\
T_4: && 
p_4'=p_4+3\varepsilon, \;\;\; p_i'= p_i-\varepsilon \mbox{ for } i \neq 4,
\label{tdef}
\eear
where $\varepsilon$ can be chosen arbitrarily small. All the transformations 
increase $\alpha_{12}$, but all have to keep the optimality of the 
probability distribution, so that minimal $\alpha$ given by $\PP'$ 
cannot be larger than $\alpha_{12}$. From transformation $T_1$ 
it follows that $\PP_0$ is optimal if and only if 
$\alpha_{12}=\min\{\alpha_{23},\alpha_{24},\alpha_{34}\}$, 
as these entries decrease under the transformation. Similarly, 
it follows from $T_2$ that $\alpha_{12}=\min\{\alpha_{13},\alpha_{14},\alpha_{34}\},$ 
and from $T_3$ that $\alpha_{12}=\min\{\alpha_{13},\alpha_{23},\alpha_{34}\},$ 
and finally from $T_4$ that $\alpha_{12}=\min\{\alpha_{14},\alpha_{24},\alpha_{34}\}$. 
Given this finite number of possibilities, one can directly check that 
either $\alpha_{12}=\alpha_{34}$ or 
$\alpha_{12}= \alpha_{13}=\alpha_{24}$ or $\alpha_{12}= \alpha_{14}=\alpha_{23}$ 
must hold for optimal probability distribution $\PP_0$ which proves the first claim.

{From} these conditions we see that there are the three candidates for the optimal $\alpha^0$:
\bea
\alpha^0&=\alpha_{12}=\alpha_{34},
\nonumber \\
&\Rightarrow 
\alpha^0=\fa^I=b_{12}+b_{34}=1 - 2\sqrt{\lambda_1\lambda_2}-2\sqrt{\lambda_3\lambda_4},
\nonumber
\\
\alpha^0&=\alpha_{13}=\alpha_{24},
\nonumber \\
&\Rightarrow 
\alpha^0=\fa^{II}=b_{13}+b_{24}=1 - 2\sqrt{\lambda_1\lambda_3}-2\sqrt{\lambda_2\lambda_4},
\nonumber
\\
\alpha^0&=\alpha_{14}=\alpha_{23}, 
\nonumber \\
&\Rightarrow 
\alpha^0=\fa^{III}=b_{14}+b_{23}=1 - 2\sqrt{\lambda_1\lambda_4}-2\sqrt{\lambda_2\lambda_3}.
\label{fadefinition}
\eea

{\it Step 2.}
At this point, we have identified three candidates for 
the $\alpha^0,$ but is is not clear yet, which one should 
be taken. 

We will show now, however, that only the minimum of these
can give a valid solution. For that, assume that one has a 
probability distribution $\PP_1$ which has the optimal 
$\alpha^0(\PP_1)= \fa^I.$ Then $\alpha_{34}^0=\alpha_{12}^0=\min_{ij}\{\alpha_{ij}\}$
and hence
\bea
\alpha_{12} \leq \alpha_{13} & \Rightarrow  b_{12}(p_1+p_3) \leq b_{13}(p_1+p_2),
\nonumber
\\
\alpha_{34} \leq \alpha_{23} & \Rightarrow  b_{34}(p_2+p_4)\leq b_{24}(p_3+p_4).
\eea
Consequently, $b_{12}+b_{34}\leq b_{13}+b_{24}$ and hence 
$\fa^{I}\leq\fa^{II}.$ Similarly, it follows that $\fa^{I}\leq\fa^{III}.$
So if one finds a solution, then it has to be the minimum of all $\fa^{k}.$

This also shows that if there is a second solution $\PP_2$ with 
$\alpha^0(\PP_2)= \fa^{II},$ then $\alpha^0(\PP_1)= \alpha^0(\PP_2)$ 
must hold, since $\fa^{I}\leq\fa^{II}$ and $\fa^{II}\leq\fa^{I}.$
Note also that the arguments leading to this did not require the assumption 
that the probability distributions have nonzero elements.

Summarizing Step 1 and Step 2, we can state that if there 
is a optimal probability distribution with non-zero elements, 
then the solution is given by 
\be
\alpha^0=\min\{\fa^{I},\fa^{II},\fa^{III}\}.
\ee

{\it Step 3.} Now we have to consider the cases where the optimal
probability distribution has some zero elements. Let us first
consider the case that there is exactly one zero element. 

There exist two possibilities. The first one arises, when the minimum is given by $\alpha_{12}$ and $p_4=0$. Then, the transformations $T_1, T_2$ and $T_4$ in Eq.~(\ref{tdef}) can still be applied, but we have to modify $T_3$, since there are no negative probabilities
\be
\hat{T}_3: 
p_3'=p_3+2\varepsilon, \;\;\; p_i'= p_i-\varepsilon \mbox{ for } i = 1,2, \;\;\; p_4'= p_4=0.
\ee
This transformation leads exactly to the same condition as $T_3$ above $\alpha_{12}=\min\{\alpha_{13},\alpha_{23},\alpha_{34}\}$. Therefore, the same conclusion as in Step 1 can be drawn. Similarly, by considering $\hat{T}_4$, one can show that if $p_3=0$, the conclusion from Step 1 still holds.

The second possibility arises, if the minimum is again given by $\alpha_{12}$, but this time $p_1=0$. Then, only $T_2$ in Eq.~(\ref{tdef}) can be applied. We define the modified transformations:
\bea
\tilde{T}_3:& 
p_3'=p_3+2\varepsilon, \;\;\; p_i'= p_i-\varepsilon \mbox{ for } i = 2,4, \;\;\; p_1'= p_1;
\nonumber
\\
\tilde{T}_4:& 
p_4'=p_4+2\varepsilon, \;\;\; p_i'= p_i-\varepsilon \mbox{ for } i = 2,3, \;\;\; p_1'= p_1;
\eea
Then, repeating the argumentation from Step 1, one arrives at the same conclusion, apart from the special case: 
$
\alpha_{12} = \alpha_{13} =  \alpha_{14} < \alpha_{kl} 
$
for $k,l \in \{2,3,4\}$ holds.

In this special case, we have that $\alpha^0=(b_{12}+b_{13}+b_{14})$
and consequently $p_k=b_{1k}/(b_{12}+b_{13}+b_{14})$ for $k=2,3,4.$
Since $\alpha_{23}= b_{23}/(p_2 +p_3) > \alpha^0=(b_{12}+b_{13}+b_{14})$ 
it follows that $b_{23}> b_{12}+b_{13}$. Generally we have
$b_{kl}> b_{1k}+b_{1l},$ for $k,l \in \{2,3,4\}$.

Due to the definition of the $b_{ij},$ it means that the Schmidt coefficients
have to fulfill
\be 
(\sqrt{\lambda_k}-\sqrt{\lambda_l})^2 >(\sqrt{\lambda_1}-\sqrt{\lambda_k})^2+(\sqrt{\lambda_1}-\sqrt{\lambda_l})^2
\ee
for $k,l \in \{2,3,4\}$. Since the $\sqrt{\lambda_k}$ are positive real numbers, 
this can only hold if $\sqrt{\lambda_1}$ is inside the interval 
$[\sqrt{\lambda_k}; \sqrt{\lambda_l}]$. As there are three intervals, and 
two of them intersect in only one point, we must have that 
$\sqrt{\lambda_1}=\sqrt{\lambda_i}$ for some $i\in \{2,3,4\}$,
which implies that the corresponding $b_{1i}=0$ and $\alpha_{1i}=0$. Since $\alpha_{12} = \alpha_{13} =  \alpha_{14}$ all of them must be zero and hence $\alpha^0=0$ and all $b_{1k}=0$ for any $k\in \{2,3,4\}$. Physically, this means that all Schmidt coefficients are the same and the state is a maximally entangled one. But then also $\fa^{I}=\fa^{II}=\fa^{III}=0,$ so this special case does not deliver a novel solution.

{\it Step 4.} Let us now consider the case, where two or more $p_i$ equal zero. 

Let us first assume that exactly two $p_i$ are zero, namely $p_2=p_3=0$. 
Then $\alpha_{14}=b_{14}$  and $\alpha_{23}=\infty$ are independent of the 
probability distribution. However, if we make the transformation
\be
\mathfrak{T}:
p_i' = p_i-\varepsilon \;\; i \in \{1,4\},
\;
\;
p_k' = p_k+\varepsilon \;\; k \in \{2,3\},
\label{tgut}
\ee
the minimal value $\alpha^0$ does not decrease (as all 
$\alpha_{12},\alpha_{13},\alpha_{24},\alpha_{34}$ remain 
constant and $\alpha_{14}$ increases). Therefore we arrive at 
a solution, where none of the $p_i$ is zero and which is as good as a solution with $p_2=p_3=0$. Thus we conclude that solutions given by distributions with two zero elements are contained in solutions characterized in Step 1.

Finally, we have to discuss the case that three $p_i$ equal zero
and consequently the remaining one equals one. This can be 
excluded with a similar transformation as in Eq.~(\ref{tgut}) 
and we leave the details as an exercise to the reader.
\qed



\end{document}